# A CRITICAL VALUE FOR DARK ENERGY

## Peter Rowlands

*Department of Physics, University of Liverpool, Oliver Lodge Laboratory, Oxford Street, Liverpool, L69 7ZE, UK. e-mail: p.rowlands@liverpool.ac.uk*

## ABSTRACT

Experimental evidence over a number of recent years has shown the density parameter of the universe $\Omega$ converging to the critical value of 1, which defines a flat, Euclidean universe. No such calculations have defined a critical value for the most significant component of $\Omega$, that for the dark energy, $\Omega_\Lambda$, but the new data provided by the Planck probe open up the previously unconsidered possibility that a particular value with special physical significance occurs at $\Omega_\Lambda = 2/3$. If future observations should converge on exactly this value, then we may have the first indication that the explanation for this phenomenon lies in necessary constraints provided by fundamental laws of physics on possible cosmologies for the universe.

The dark energy content of the universe, discovered in 1998 (Schmidt et al. 1998, Perlmutter et al. 1999), and generally described as completely unexpected and without obvious explanation, is a phenomenon with significant implications for physics as well as cosmology. In fact, the new results available from the Planck probe allow the previously unexpected possibility that the dark energy may show fundamental physics driving possible cosmologies, rather than cosmology determining possible physics, and this can be done directly from the data without requiring any conjectural or speculative physics input. In addition, we can immediately see how such a possibility can be put to rigorous testing using data from future probes.

In relation to the scale factor $R$, in an expanding universe, the Hubble parameter $H$ is defined as the normalised rate of expansion $H = \dot{R}/R$, and is measured as the ratio $v/r$ of the recessional velocity $v$ and the comoving distance, $r$, of distant galaxies. The Hubble constant, $H_0$, the Hubble parameter at the present time, is defined in terms of the Hubble radius, $r_H$, as $c/r_H$. So, at the present time,

$$v = H_0 r = \frac{cr}{r_H}. \qquad (1)$$

Friedmann's solutions of the Einstein field equations (Friedmann 1922) suggest that there is a particular density at which the universe must be flat or Euclidean, with curvature parameter $k$ equal to zero. According to the first Friedmann equation, in the absence of a cosmological constant $\Lambda$,

$$H^2 = \left(\frac{\dot{R}}{R}\right)^2 = \frac{8\pi G}{3}\rho - \frac{kc^2}{R^2}. \tag{2}$$

At the present time, the critical density for zero curvature is given by

$$\rho_{crit} = \frac{3H_0^2}{8\pi G} \tag{3}$$

and the ratio of the actual density $\rho$ to this critical value, the density parameter, $\Omega = \rho / \rho_{crit}$, will determine the universe's evolution and ultimate fate. The value $\Omega = 1$ clearly has a special significance, creating a clear separation of a flat or Euclidean universe from a closed universe with $\Omega > 1$ and spherical geometry, and an open universe with $\Omega < -1$ and hyperbolic geometry. The fact that, since 2000 (Bernardis et al. 2000), the experimental value of this parameter has been apparently converging towards the precise value of 1, has been taken as indicating that the universe is flat and possibly infinite, apparently in line with an inflationary view of cosmology, but in contradiction with the consensus that became accepted during the main part of the twentieth century for the geometry of the universe. It also means that general relativistic calculations are taken at the Newtonian limit, if we incorporate energy into the mass term. Then equation (3) becomes equivalent to $\rho_{crit} = 3\,v^2 / 8\,\pi G r^2$, which implies the Newtonian relation $Gm / r = \tfrac{1}{2}\,v^2$, where $m$ is the mass enclosed within radius $r$.

The particular form of the Friedmann equation in equation (2) allows the possibility of $\rho$ being replaced by $\rho + 3P / c^2$, and the inclusion of a vacuum energy term with positive (inward) pressure $P$ from, say, radiation, without any significant change to the meaning of $\Omega$. However, the inclusion of a 'cosmological constant' $\Lambda$ or a vacuum energy term with negative (outward) pressure, as in equation (4), while, not changing the definition of $\Omega$, would have major physical consequences.

$$H^2 = \left(\frac{\dot{R}}{R}\right)^2 = \frac{8\pi G}{3}\rho - \frac{kc^2}{R^2} - \frac{\Lambda c^2}{3} \tag{4}$$

It was a term of this kind that was discovered in 1998, through an outward acceleration in the red-shift velocity of distant galaxies (Schmidt et al. 1998, Perlmutter et al. 1999). Named 'dark energy', and incorporated into $\Omega$ as the component $\Omega_\Lambda$, it has remained an unexplained phenomenon, and the early values of its magnitude, ranging from 71.4 to 74 per cent of the total energy of the universe, gave no obvious clue as to its origin. However, it may be that, just as with $\Omega$ itself, there is a particular critical value of $\Omega_\Lambda$ with a precise physical significance, and it may be that the new data from the Planck probe indicate that it could converge towards this particular value.

The main result quoted for $\Omega_\Lambda$, from Planck Collaboration XVI, gives 0.6825 as the best fit and 0.686 ± 0.020 for the 68 % confidence limits (Planck Collaboration XVI 2013). Including lensing as well as Planck gives 0.6964 as best fit and 0.693 ± 0.019 for the 68 % confidence limits. Including WMap as well as Planck gives 0.6817 as best fit and 0.685 +

0.018 and − 0.016 for the 68 % confidence limits. The overview reported by Planck Collaboration I states that determining the dark energy contribution from temperature anisotropies data alone gives 0.67 + 0.027 and − 0.023 for the 68 % confidence limits (Planck Collaboration I 2013).

The new value for $\Omega_\Lambda$ suggests an intriguing possibility. The value for this vacuum energy is close to two-thirds of the total energy of the universe, and, if this fraction should turn out to be the preferred value, then a simple calculation suggests some interesting consequences. If we suppose that

$$\frac{\rho_{vac}}{\rho_{crit}} = \frac{2}{3}, \tag{5}$$

then, using (3), the vacuum density becomes

$$\rho_{vac} = \frac{H_0^2}{4\pi G}. \tag{6}$$

This is equivalent to a 'dark' energy density or negative pressure

$$-P = \frac{H_0^2 c^2}{4\pi G}, \tag{7}$$

and cosmological constant

$$\Lambda = 8\pi G \rho_{vac} = 2H_0^2. \tag{8}$$

We can incorporate $P$ (or $\Lambda$) into equation (4), with $k = 0$, to give

$$H^2 = \left(\frac{\dot{R}}{R}\right)^2 = \frac{8\pi G}{3}\left(\rho + \frac{3P}{c^2}\right) \tag{9}$$

or into Friedmann's second, acceleration, equation, to give

$$\frac{\ddot{R}}{R} = \frac{4\pi G}{3}\left(\rho + \frac{3P}{c^2}\right). \tag{10}$$

In line with our observation that this is at the Newtonian limit, we can see the connection between these equations and an equivalent Poisson equation:

$$\nabla^2 \phi = 4\pi G\left(\rho + \frac{3P}{c^2}\right) = 4\pi G\left(\rho - \frac{3H_0^2}{4\pi G}\right) = 4\pi G(\rho - 3\rho_{vac}). \tag{11}$$

If $\rho$ is the mass density of a uniform and isotropic Hubble universe with mass $m = 4\pi G \rho r^3 / 3$, within radius $r$, then we can express equation (10) in terms of a force on a unit mass, combining the effect of gravity and dark energy:

$$F = \frac{Gm}{r^2} - H_0^2 r = \left(\frac{4}{3}\pi G\rho - H_0^2\right)r. \tag{12}$$

This means that the acceleration responsible for dark energy can be expressed as

$$a = \frac{v^2}{r_H} = H_0^2 r. \tag{13}$$

This is a remarkable result, suggesting that the acceleration observed in the red-shift, like the velocity, depends only on Hubble's constant $H_0$ and the distance, and is thus perhaps an integral component of the same process that produces the red-shift velocity $v$. Integrating

$$a = v\frac{dv}{dr} = H_0^2 r \tag{14}$$

with respect to $v$ and $r$ between the limits 0 and $v$ and 0 and $r$ gives the exact Hubble red-shift law:

$$v = H_0 r = \frac{cr}{r_H}. \tag{15}$$

Here, we should point out that the significant equation (14) has been obtained solely from the data, with no conjectural or speculative element whatsoever, and no additional theoretical content or modelling. If the data stands as it is today, the equation is valid to within the experimental confidence limits, and certainly to within a factor $1.02 \pm 0.02$. It is so close to being exactly true, and of such exceptional physical significance if it is, that we are justified in saying that the test of its exact validity should be one of the aims of future probes. We may, for comparison, recall that the Boomerang collaboration (Bernardis et al. 2000), finding the 95% confidence interval for $\Omega$ to be between 0.88 and 1.12, were immediately able to claim, with full justification, that this provided 'evidence for a euclidean geometry of the Universe'. Even now, $\Omega = 1$ is hardly established to better than about 1 %, erring on the side of an increased $\Omega$ value, and inferentially that of $\Omega_\Lambda$. Planck Collaboration I, including data from lensing, constrains 'departures from spatial flatness at the percent level', that is $\Omega_k = -0.0096$ with 68 % confidence limits of $+ 0.010 - 0.0082$, that is, a total $\Omega$ of $1.0096 + 0.0082 - 0.010$ (Planck Collaboration I 2013).

In the present case, the physical significance of equation (14) stems from the fact that the velocity term can be derived directly from the acceleration, implying that the acceleration, whatever its origin, is actually *responsible* for the velocity. Clearly, if this is true, there must be a significant impact on possible models of cosmological evolution. There is, however, another equation which we can derive directly from (5) and (14), which, again without any hypothetical or model-dependent input, suggests even further significance. We begin by writing the acceleration in the form

$$a = \frac{dv}{dt} = \frac{c^2 r}{r_H^2}, \tag{16}$$

following which we recall that Sciama considered the possibility of explaining inertia along the lines of Mach's principle using a gravitomagnetic inductive force between two masses with relative acceleration, which could be derived from general relativity (Sciama 1953, 1972). In this case, there is an inductive force between masses $m_1$ and $m_2$,

$$F = \frac{G}{c^2 r} m_1 m_2 \sin\theta \frac{dv}{dt} \tag{17}$$

of the same kind as the one between charges $e_1$ and $e_2$,

$$F = \frac{G}{c^2 r} e_1 e_2 \sin\theta \frac{dv}{dt}, \tag{18}$$

which can be derived from Faraday's law of induction. Sciama considered that, using the gravitomagnetic inductive force, and assuming that isotropy removes the angular dependence $\theta$, the inertia of a body of mass $m = m_1$ could be attributed to the action of the total mass $m_H = m_2$ within the observable universe, specified by radius $r_H$, so making the inductive force equation equivalent to the Newtonian inertial equation $F = Kma$, with $K$ a constant and $a = dv / dt$. The inertial force on a unit mass due to the entire mass in the Hubble universe $m_H$ would then be:

$$F = \frac{G}{c^2 r} m_H \frac{dv}{dt}. \tag{19}$$

Inertia provides one standard for defining a unit mass, gravitation provides another, and the connection can be made via the equivalence principle. If we now suppose that mass $m_H$ defines a radial inertial field of constant magnitude from the centre of a local coordinate system, and, at the same time, use the principle of equivalence to equate the magnitude of this to the gravitational field ($Gm_H / r_H^2$), which, independently of the local coordinate system, defines a unit of gravitational mass within the same event horizon, we obtain

$$\frac{Gm_H}{c^2 r} \frac{dv}{dt} = \frac{Gm_H}{r_H^2}, \tag{20}$$

which gives us the exact expression for the acceleration which would result if the dark energy constitutes exactly two-thirds of the total energy of the universe:

$$a = \frac{c^2 r}{r_H^2} = H_0^2 r. \tag{21}$$

The calculation suggests that an exact value of two thirds for the dark energy contribution would not only link the dark energy and the Hubble red-shift as aspects of the same phenomenon, but could be of additional interest in connection with Mach's principle and the origin of inertia. Again, there is an indication that possible cosmologies are necessarily constrained by fundamental laws of physics. A Machian origin of inertia, for example, would allow a universe to evolve by creating inertial mass at the same time as its space-time structure, with the creation process also generating the force which drives its evolution. It is significant, of course, that equation (20) is simply another version of equation (16), and does not *require* the development through equations (17)-(19) for its derivation. These serve to provide a *possible* context, but none is needed to generate the equation, and the form of the equation is in itself significant. However, prior prediction leading to experimental confirmation remains one of the strongest arguments available for any theoretical construction, and, in the present case, there is also a prior prediction.

A version of the calculation was done in reverse as part of a larger study when values of $\Omega_\Lambda$ looked less favourable to its conclusions (Rowlands 2007), and this was preceded by a series of calculations deriving the red-shift acceleration as $H_0^2 r$ on the basis of a flat universe, some of which predated the experimental discovery of the dark energy. The most accessible, though not the earliest version of $a = H_0^2 r$, from a series of publications beginning in 1979, was incorporated into a book with a largely historical slant (Rowlands, 1994). In this predictive context, the dark energy would seem to have a possible explanation in both physical and cosmological contexts. The calculations seem to imply that there may be a critical value for $\Omega_\Lambda$ just as there is for $\Omega$. This potentially critical value now falls within the limits of the data provided by the Planck probe, and future experimental findings may converge on this value, just as they have converged on the physically significant value of unity for $\Omega$. Even a value which came very close would require explanation in the same way as values of $\Omega$ close to 1 were thought to be too close for coincidence even before observations were able to establish an exact value. It would be interesting to see how the constraints on other cosmological parameters would be affected by applying an exact value of two thirds for the dark energy density to the Planck data, and how any possible deviations in the assumed universal isotropy and uniformity might be manifested.